\documentstyle[12pt]{article}

\begin{document}
\baselineskip=24pt
\thispagestyle{empty}
\title{
Necessary and Sufficient Condition for Nonsingular
Fisher Information Matrix in ARMA Models\/}
\author{A. Ian McLeod \\ Department of Statistical and Actuarial Sciences
\\ University of Western Ontario \\ London, Ontario N6A 5B7
\\ Canada }
\date{}
\maketitle

\bigskip
\begin{center}
{\bf Abstract}
\end{center}
It is demonstrated that a necessary and sufficient condition that
the Fisher information matrix of an ARMA model be nonsingular
is that the model not be redundant, that is
the autoregressive and moving-average polynomials do not share common roots.

\vspace*{.5in}

\vfill\eject

\bigskip\par

The ARMA($p,q$) model, which may be written in operator notation as,
$\phi(B) z_t = \theta(B) a_t,$
where $\phi(B) = 1 - \phi_1 B - \ldots - \phi_p B^p, \theta(B) = 1 - \theta_1 B - \ldots -
\theta_q B^q, a_t $ is white noise with variance $\sigma_a^2>0 $
and $B$ is the backshift operator on $t$, is said to be
not redundant if and only if $\phi(B)=0$ and $\theta(B) = 0$ have no common
roots.

The Fisher information
matrix may be written (Box and Jenkins, 1976, p.240),
$ I(\phi, \theta) = \sigma_a^{-2} {\rm E}\{A_t A_t^{\prime}\}$,
where
$A_t = (v_{t-1}, \ldots, v_{t-p}, u_{t-1}, \ldots, u_{t-q})$,\
$\phi(B)v_t = -a_t$,\   $\theta(B) u_t = a_t$ and
$A_t^{\prime}$ denotes the transpose of $A_t$.

The inverse of this Fisher information matrix divided by the length
of an observed time series can be used to obtain an estimate
of the covariance matrix of the estimated parameters $\phi_1, \ldots,
\phi_p, \theta_1, \ldots, \theta_q$.
McLeod (1984) and Klein and M\'elard (1990) have given algorithms
for computing the inverse of the Fisher information matrix.
It is frequently assumed without proof in the time series literature
that the Fisher information
matrix is nonsingular provided the ARMA model is not redundant.
For examples see,
Brockwell and Davis (1991, Theorem 8.11.1),
and Hannan (1970, Theorem 8, p.403 and pp.413--414).
Poskitt and Tremayne (1981, p.977) have proved this
result but their approach is more complicated than that given below.

{\it Theorem\/}: The information matrix $I(\phi, \theta)$
is nonsingular if and only if the model is not redundant.

The proof of this theorem uses the following algebraic lemma.

\medskip
{\it Lemma\/}: A necessary and sufficient condition for model redundancy is
that there exist nonzero polynomials
$\alpha(B) = \prod_{j=1}^{p-1}(1-E_j B),$
and
$\beta(B) = \prod_{j=1}^{q-1}(1-F_j B),$
such that
$$\alpha(B) \theta(B) = \beta(B) \phi(B). \eqno(1.1)$$
\medskip
{\it Proof of Lemma \/}: Let
$\phi(B) = \prod_{j=1}^p (1-G_i B),$
and
$\theta(B) = \prod_{j=1}^q (1-H_i B).$
Then (1.1) can be written,
$$\prod_{i=1}^{p-1} (1-E_i B) \prod_{j=1}^q (1-H_j B) =
\prod_{i=1}^{q-1} (1-F_i B) \prod_{j=1}^p (1-G_j B).  \eqno(1.2)$$
If $H_1 = G_1$ then (1.2) holds by setting
$E_i = G_{i+1},\ (i=1,\ldots,p-1)$ and
$F_i = H_{i+1},\ (i=1,\ldots,q-1)$.
Conversely, if $G_i \ne H_j,\ \ \forall i, j$ then $\phi(B) / \theta(B)$
is irreducible and must be distinct from $\alpha(B) / \beta(B)$ since
the degree of the polynomials occurring in the numerator and denominator
is $p$ and $q$ in the latter irreducible case and
$p-1$ and $q-1$ in the former.
\medskip

{\it Proof of Theorem\/}:
$I(\phi, \theta)$ is singular if and only if there exist
$\alpha_0, \ldots, \alpha_{p-1}$ and $\beta_0, \ldots, \beta_{q-1}$,
not all zero, such that
$${\rm Var}(\sum_{i=0}^{p-1} \alpha_i v_{t-i} + \sum_{j=0}^{q-1} \beta_j u_{t-j}) = 0
\eqno(1.3) $$
where $\alpha_i \ne 0$ for some $i$ and
$\beta_j\ne 0$ for some $j$.
This follows because
$ I(\phi, \theta) = \sigma_a^{-2} {\rm E}\{A_t A_t^{\prime}\}$
is nonnegative definite.
Eq. (1.3)  is equivalent to the condition that there exists
polynomials
$\alpha(B) = \alpha_0 + \alpha_1 B + \ldots + \alpha_{p-1} B^{p-1}$
 and
$\beta(B) = \beta_0 + \beta_1 B + \ldots + \beta_{q-1} B^{q-1}$,
at least one of which is nonzero,
such that
$$\alpha(B) v_{t-1} + \beta(B) u_{t-1} = 0. \eqno(1.4)$$
Eq. (1.4) can now be written
$${\alpha(B)\over \phi(B)}a_t - {\beta(B)\over \theta(B)} a_t = 0.
\eqno(1.5)$$
Multiplying (1.5) by $a_t$ and taking expected values, we see
that
$\alpha(B)/\phi(B) - \beta(B)/\theta(B) = 0$
which by the Lemma is a necessary and sufficient condition for model
redundancy.
This establishes that lack of model redundancy is a necessary and
sufficient condition for the Fisher information matrix, $I(\phi, \theta)$
to be nonsingular.

\medskip
\begin{center}
{\bf Acknowledgments}
\end{center}
Improvements suggested by two anonymous referees are acknowledged.

\bigskip

\vspace*{0.5cm}
\def\bib{\noindent\hangindent=15pt}

\begin{center}
{\bf REFERENCES}
\end{center}
\bigskip
\bib BROCKWELL, P.J. and DAVIS, R.A. (1991),
{\it Time Series: Theory and Methods\/},
(2nd edn), New York: Springer-Verlag.

\medskip
\bib BOX, G.E.P. and JENKINS, G.M. (1976),
{\it Time Series Analysis: Forecasting and Control\/},
(2nd edn), San Francisco: Holden-Day.

\medskip
\bib HANNAN, E.J. (1970),
{\it Multiple Time Series\/},
New York: Wiley.

\medskip
\bib MCLEOD, A.I. (1984),
Duality and other properties of multiplicative seasonal
autoregressive-moving average time series models,
{\it Biometrika\/} 71, 207--211.

\medskip
\bib KLEIN, A. AND M\'ELARD, G. (1990),
Fisher's information matrix for seasonal autoregressive-moving average
models,
{\it Journal of Time Series Analysis\/} 11, 231--237.

\medskip
\bib POSKITT, D.S. and TREMAYNE, A.R. (1981),
An approach to testing linear time series models,
{\it The Annals of Statistics\/} 9, 974--986.

\end{document}